\documentclass{elsart}
\usepackage{graphicx}
\usepackage{amssymb}
\usepackage{amsmath}
\begin{document}

\begin{frontmatter}
\title{Exact Enumeration of Three--Dimensional Lattice Proteins}

\author{Reinhard Schiemann}\ead{Reinhard.Schiemann@itp.uni-leipzig.de}, 
\author{Michael Bachmann}\ead{Michael.Bachmann@itp.uni-leipzig.de}, and
\author{Wolfhard Janke}\ead{Wolfhard.Janke@itp.uni-leipzig.de}

\address{Institut f\"ur Theoretische Physik, Universit\"at Leipzig,\\Augustusplatz 10/11, 
         04109 Leipzig, Germany}

\begin{abstract}
We present an algorithm for the exhaustive enumeration of all monomer
sequences and conformations of short lattice proteins as described by
the hydrophobic--polar (HP) model. 
The algorithm is used for an exact identification of all designing sequences
of HP proteins consisting of up to 19 monomers whose conformations are
represented by interacting self--avoiding walks on the simple cubic lattice. 
Employing a parallelized implementation on a Linux cluster, we
generate the complete set of contact maps of such walks. 
\end{abstract}


\end{frontmatter}

\section{Introduction}
The numerical treatment of protein models is highly nontrivial. On one hand, the design
of realistic models suffers from the fact that the atomic interactions among the
constituents of proteins
and with their aqueous cellular environment are by no means well understood~\cite{Crei93}. 
On the other
hand, the computational effort increases drastically with the length of the molecules.
Therefore, significant simplifications of the realistic situation have to be introduced in order to 
facilitate a detailed analysis based on computational methods and, in particular, to allow
studies of the relation between sequence and conformation spaces of model proteins.

Herein, we will consider two versions
of the very simple HP lattice model~\cite{Dill85,Tang00} which makes the following assumptions:
Instead of considering all 20 different kinds of amino acids that occur in real proteins
the model comprises only two prototypes of residues: hydrophilic (or polar, $P$) and
hydrophobic ($H$) monomers, respectively. This is
to account for the fact that most of the naturally occurring amino acids can be
classified in that way (Ref.~\cite{Crei93}, p.~154). Also the atomar interactions
are drastically simplified.
Short-range repulsion between monomers is taken into account by modeling the
conformations of HP proteins as \emph{self--avoiding walks} on regular lattices. 
The simple cubic (sc) lattice was used in this study.
In addition one considers in the most simple formulation of the model exclusively 
a nearest--neighbor attractive interaction between 
hydrophobic residues non--adjacent in the polymer chain~\cite{Dill85}. Slightly more involved
variants also take into account nearest-neighbor contacts between $HP$ and/or $PP$ 
pairs~\cite{Tang00}.
This is an effective way of describing the interaction of the molecule with the aqueous
environment~\cite{Miya96}.

Exact enumeration results obtained for short HP proteins can be used as cross--checks for
other non--exact methods that search the conformational and sequence spaces
of proteins. These include Monte Carlo and genetic algorithms (e.g.\ Refs.~\cite{Soka95,Ung93}),
generalized ensemble techniques (e.g.\ Ref.~\cite{Okam98}), chain growth
algorithms (e.g.\ Refs.~\cite{Born97,Gras97}), and combinations
thereof (e.g.\ Refs.~\cite{Mit01,Bach03}).
More importantly, the complete treatment of all sequences and conformations allows
one to carry out systematic statistical analyses of HP proteins.
Our results for the sc lattice described in more detail in Ref.~\cite{Schi04} complement prior exact enumeration
studies on the square lattice\,\cite{Irb02} and for HP proteins with conformations
{\em restricted} to regular cuboids on the sc lattice\,\cite{Li96,Cejt02}.

In the next section we introduce the HP models used here in a little more
formal way. Section~\ref{sec_EE} explains the exact enumeration procedure in terms
of which our results are obtained. The concept of exact enumeration is first illustrated
with a naive implementation. What remains of Section~\ref{sec_EE} is
dedicated to improvements of that simple implementation and describes how these
improvements apply to a simple example case. In Section~\ref{sec_appl} we show
how our exact results can be applied for a comparison of the numbers
of self--avoiding walks and contact matrices and for the determination
of designing sequences in the HP model. Section~\ref{sec_sum} concludes this article with
a summary and an outlook on further statistical analyses based on the
results of the enumerations presented here.

\section{HP Models}
An HP protein is defined by its sequence of monomers. 
We will denote the type of monomer by $\sigma_i$, where
$i=1,\ldots,N$ is the position of the monomer in a polymer chain of length $N$
and by convention $\sigma_i \in \lbrace 0\widehat{=} P,1\widehat{=} H\rbrace$.
Its conformation, which is a self--avoiding walk on the lattice (with lattice spacing $a=1$), 
is represented by an
ordered collection of lattice vectors that contain the positions of the
residues:\ $\mathbf{X}=(\mathbf{x}_1,\mathbf{x}_2,\ldots,\mathbf{x}_N)$. 
The distance between monomers $i$ and
$j$ is denoted by $x_{ij}=|\mathbf{x}_i - \mathbf{x}_j|$.  
The attractive interaction between pairs of residues is short--ranged on the underlying
lattice. It is considered only between residues that are on nearest-neighbor
positions but not covalently bound in the molecular chain. Such a pair of residues
is said to be in \emph{contact}. This is expressed by the
following energy function that is assigned to each HP protein:
\begin{equation}
 \label{eqn_HP_Emain}
 E=\sum_{i,j>i+1}C_{ij}\,U_{\sigma_i \sigma_j},
\end{equation}
where $C_{ij}=(1-\delta_{i+1 j})$ for $x_{ij}=1$, and zero otherwise, 
is a symmetric $N\times N$ matrix called \emph{contact map} and
\begin{equation}
 U_{\sigma_i \sigma_j}=
 \left(
  \begin{array}{cc}
   u_{HH} & u_{HP}\\
   u_{HP} & u_{PP}
  \end{array}
 \right)
\end{equation}
is the $2\times 2$ interaction matrix.

In the present study, the HP model comes in two versions that
are different from one another in the way attractive interactions between the
amino acids are considered. In the original version of the model\,\cite{Dill85},
which we will refer to as HP model in the following, only a pair of hydrophobic
residues in contact contributes to the energy function (\ref{eqn_HP_Emain}) and the
only non--zero entry in the interaction matrix is $u^\text{HP}_{HH}=-1$.
A modification of the model\,\cite{Tang00} also takes into account an
interaction between hydrophobic and polar monomers. We call it the 
MHP (mixed HP) model. Its interaction matrix entries read $u^\text{MHP}_{HH}=-1$, 
$u^\text{MHP}_{HP}=-1/2.3 \approx -0.435$, and $u^\text{MHP}_{PP}=0$.
The magnitude of $u^\text{MHP}_{HP}$ is motivated by an analysis of
inter--residue contact energies between different types of real amino
acids\,\cite{Miya96}.

A sequence of monomers is called a \emph{designing sequence} if there exists
exactly one conformation (up to trivial symmetries, to be explained in more detail
below) for its state of lowest energy. The interest in designing sequences
is based on a generally accepted biochemical principle that
sequence specifies conformation, and, in turn, the conformation of a polymer
determines its biological function. Accepting this principle
also in the framework of the highly simplified HP model leads directly to the
concept of designing sequences. The conformation of the lowest--energy state
is uniquely determined for designing sequences only. Furthermore,
the number of designing sequences is very small compared to
the total number of $2^N$ HP sequences of a given chain length $N$. Thus, the ability
of identifying designing sequences may be seen as a benchmark for
algorithms that search the sets of conformations {\em and} sequences
of HP proteins.

\section{Exact Enumeration}
\label{sec_EE}
\subsection{Naive implementation}
A straightforward method of identifying designing sequences of a given length
is to perform an exact enumeration. This means to run through
the whole set of sequences and for each sequence through the whole set of
conformations. Consider such a deliberately naive enumeration for short
sequences of length $N=4$ in the HP model. 
Trivially, there are $2^4=16$ different sequences
and $6\times 5\times 5=150$ self--avoiding walks on the sc lattice.

Up to symmetries, Fig.~\ref{fig_sawN4} shows all conformations for $N=4$.
Only the conformation designated by \texttt{FLL} has a contact between its
first and last residues.
Consequently, all four HP sequences with a hydrophobic monomer in the
first and last positions of the sequence must be designing: there is
only one conformation for the lowest energy $E=-1$. All
other sequences are non--designing since the energy $E=0$ is obviously 
degenerate.

When increasing
the number of monomers $N$ by one, the number of sequences doubles. Also,
the number of self--avoiding walks (SAW) is known to increase asymptotically by a
factor of $\mu_\text{SAW} \approx 4.684$\,\cite{MacD92,MacD00,Chen02}. 
Consequently, the computational effort scales
roughly as $9.37^N$, i.e., exponentially fast with the chain length $N$. 
This is why improvements of the naive enumeration become
necessary even for rather short chain lengths.
\subsection{Improvements}
Some of these improvements are very obvious. Firstly, there is no point in
carrying out the enumeration for two sequences that contain the same monomers
but in reverse order. For example, the sequences \texttt{HPPP} and \texttt{PPPH}
are equivalent.
Simply counting the number of relevant sequences, $R_N$, that have to be considered
in the enumeration yields
\begin{equation}
 \label{eqn_relseqs}
 R_N = 2^{N-1} +
 \begin{cases}
  \ 2^{\frac{N}{2}-1} & \text{if $N$ even},\\
  \ 2^{\frac{N-1}{2}} & \text{if $N$ odd}.
 \end{cases}
\end{equation}
\subsubsection{Symmetries on the sc lattice}
\begin{figure}
 \begin{center}
 \begin{tabular}{clclcl}
  \includegraphics{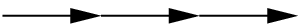} & \texttt{FFF}&
  \includegraphics{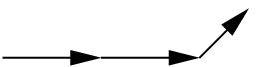} & \texttt{FFL}&
  \includegraphics{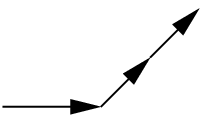} & \texttt{FLF}\\
  \includegraphics{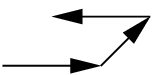} & \texttt{FLL}&
  \includegraphics{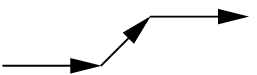} & \texttt{FLR}&
  \includegraphics{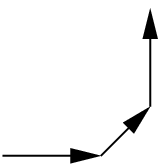} & \texttt{FLU}\\
 \end{tabular}
 \end{center}
 \caption{All relevant conformations for $N=4$ together with their configurational
chain codes explained in the text.}
 \label{fig_sawN4}
\end{figure}
\begin{figure}[b]
 \begin{center}
  \begin{tabular}{clclclcl}
   \includegraphics{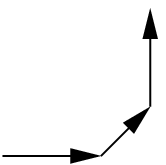} & (a) &
   \includegraphics{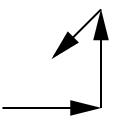} & (b) &
   \includegraphics{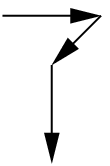} & (c) &
   \includegraphics{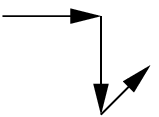} & (d) \\
   \includegraphics{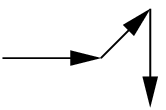} & (e) &
   \includegraphics{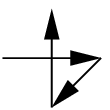} & (f) &
   \includegraphics{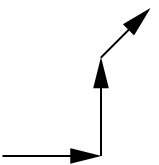} & (g) &
   \includegraphics{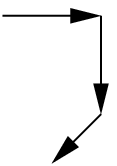} & (h) \\
  \end{tabular}
 \end{center}
 \caption{All $(1/6)\,f_\text{s}(\mathbf{X})=8$ conformations are symmetric on the sc lattice
          and their first bonds show into the same direction. Symmetry
          operations applied to the conformation (a) are rotations about the
          first bond~(b,c,d), reflections at a lattice plane (e,f), and
          two compositions of rotations and reflections~(g,h).}
 \label{fig_symm_FLU}
\end{figure}
\begin{figure}
 \begin{center}
 \includegraphics{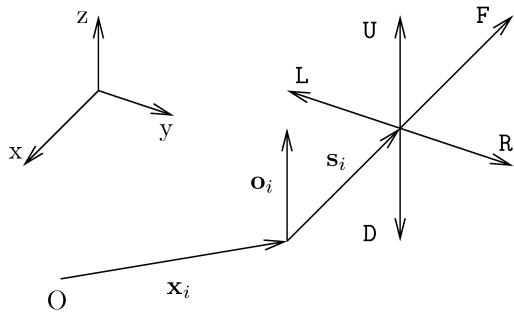}
 \end{center}
 \caption{Encoding a conformation in terms of relative moves on the
          lattice. Vectors designated by \texttt{F}, \texttt{L},
          \texttt{R}, \texttt{U}, \texttt{D} are $\mathbf{s}_{i+1}(\texttt{F})$,
          $\mathbf{s}_{i+1}(\texttt{L})$, etc.}
 \label{fig_moves}
\end{figure}

Furthermore, not all self--avoiding walks are independent but related to each
other by symmetry operations. On the sc lattice, there are six
directions for the first bond of any conformation. Fixing the direction of
the first bond, we are left with $(1/6)\,f_\text{s}(\mathbf{X})$ mutually symmetric
conformations, where $f_\text{s}(\mathbf{X})$ is the total number of mutually symmetric
self--avoiding walks starting from the origin.
Figure~\ref{fig_symm_FLU} illustrates these symmetries
for the conformation encoded by \texttt{FLU} in Fig.~\ref{fig_sawN4}.
For a given conformation $\mathbf{X}$, the symmetry factor is given by
\begin{equation}
 \label{eqn_fS}
 f_\text{s} \left( \mathbf{X} \right) =
 \begin{cases}
  \ 6 & \text{if } \mathbf{X} \text{ linear},\\
  \ 24 & \text{if } \mathbf{X} \text{ planar},\\
  \ 48 & \text{ otherwise.}
 \end{cases}
\end{equation}
We represent
conformations by means of chain codes that encode the steps of
self--avoiding walks in terms of a sequence of relative moves.
On the sc lattice there are five kinds of such moves
which we denote by $\texttt{F}$ (``forward''), $\texttt{L}$ (``left''),
$\texttt{R}$ (``right''), $\texttt{U}$ (``up''), and $\texttt{D}$ (``down'').
The chain codes for all independent conformations consisting
of four monomers are shown in Fig.~\ref{fig_sawN4}.
Two vectors are needed in order to define the five moves on the sc lattice:
Let $\mathbf{o}_i$ be a unit vector attached to the monomer at 
$\mathbf{x}_i$ and $\mathbf{s}_i$ another unit vector at $\mathbf{x}_i$
perpendicular to $\mathbf{o}_i$ determining the direction of the
$i$\textsuperscript{th} step of the self--avoiding walk as shown
in Fig.~\ref{fig_moves}.
Given a chain code, we determine the conformation $\mathbf{X}$ by initially
choosing $\mathbf{x}_1=(0,0,0)$, $\mathbf{o}_1=(0,0,1)$, $\mathbf{s}_1=(0,1,0)$
and by specifying how to go over from $\lbrace \mathbf{x}_i,\mathbf{o}_i,
\mathbf{s}_i \rbrace$ to $\lbrace \mathbf{x}_{i+1},\mathbf{o}_{i+1},
\mathbf{s}_{i+1} \rbrace$ for each of the five moves:
\begin{align}
 \label{eqn_omove}
 \mathbf{o}_{i+1}&=
 \begin{cases}
  \ \mathbf{o}_i & \text{for moves \texttt{F},\texttt{L},\texttt{R}},\\
  \ -\mathbf{s}_i & \text{for \texttt{U}},\\
  \ \mathbf{s}_i & \text{for \texttt{D}},
 \end{cases}\\
 \notag\\[-0.5cm]
 \label{eqn_smove}
 \mathbf{s}_{i+1}&=
 \begin{cases}
  \ \mathbf{s}_i & \text{for \texttt{F}},\\
  \ \mathbf{o}_i \times \mathbf{s}_i & \text{for \texttt{L}},\\
  \ -\mathbf{o}_i \times \mathbf{s}_i & \text{for \texttt{R}},\\
  \ \mathbf{o}_i & \text{for \texttt{U}},\\
  \ -\mathbf{o}_i & \text{for \texttt{D}},
 \end{cases}\\
 \notag\\[-0.5cm]
 \label{eqn_rmove}
 \mathbf{x}_{i+1} &= \mathbf{x}_i + \mathbf{s}_i.
\end{align}
Equations (\ref{eqn_omove}), (\ref{eqn_smove}), and (\ref{eqn_rmove})
can be read off from Fig.~\ref{fig_moves}.

In exact enumeration, it is not desirable to
enumerate conformations that are symmetric to each other. This is easily
achieved by enumerating the chain codes for conformations of a given
length $N$ considering only codes that satisfy a chosen set of rules. 
The choice of $\mathbf{x}_1$, $\mathbf{o}_1$, and $\mathbf{s}_1$ determines the first
move which we call by convention \texttt{F}. Furthermore, we require
the first move that makes the walk deviate from a linear conformation
to be an \texttt{L}--move and, subsequently, we require the first
step into the third coordinate direction to be a \texttt{U}--move.
For conformations of length $N=4$ modeled as self--avoiding walks of three
steps there are six different chain codes obeying these rules
and coding for not mutually symmetric conformations (see again Fig.~\ref{fig_sawN4}).
\subsubsection{Contact maps}
\label{ssec_cm}
\begin{figure}
 \begin{center}
  \includegraphics{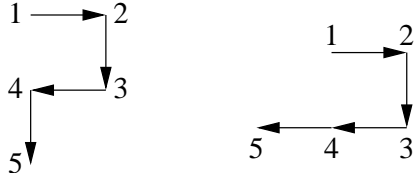}
 \end{center}
 \caption{Two different conformations which have a single contact between their first and
          fourth monomers. Both belong to the same contact map.}
 \label{fig_saw_cm}
\end{figure}

As defined in (\ref{eqn_HP_Emain}), the energy of an HP protein does not depend
explicitly on its conformation but only on the information of which pairs of
monomers form contacts. This information is contained in the contact map.
In general, more than one conformation corresponds to a given contact map
(see Fig.~\ref{fig_saw_cm}).
Therefore, it is possible to improve exact enumeration in terms of contact maps:
First, all self--avoiding walks of a given length are enumerated \emph{once} in
order to generate the complete set of contact maps. In a second step, designing
sequences are identified by running through the set of contact maps for each
sequence. A sequence is identified as designing if there is exactly one contact
map that corresponds to the lowest energy and, in turn, if there is exactly one
self--avoiding walk corresponding to that contact map.

The first step of this enumeration procedure requires all contact maps
to be stored in memory. Some straightforward properties of contact maps can
be used in order to occupy as little memory as possible.
Apart from symmetry ($C_{ij} = C_{ji}$) and the trivial facts that self-contacts are not
meaningful ($C_{ii}=0$) and that, by definition, covalently bound monomers are not
counted as contacts ($C_{ij}=0$ if $|i-j| = 1$), these properties are:
\begin{align}
 C_{ij} &= 0 \quad \text{if }|i-j|=2n, \quad n=1,2,\ldots, \label{eqn_prop3}\\
 \sum_{i=1}^{N}C_{ij} &\leq
 \begin{cases}
  \ 5 & \text{if } j = 1\ {\rm or}\ j = N,\\
  \ 4 & \text{otherwise},
 \end{cases} \label{eqn_prop4}
\end{align}
which are easily seen to be consequences of the considered sc lattice geometry.
\begin{figure}
 \begin{center}
  \includegraphics[clip]{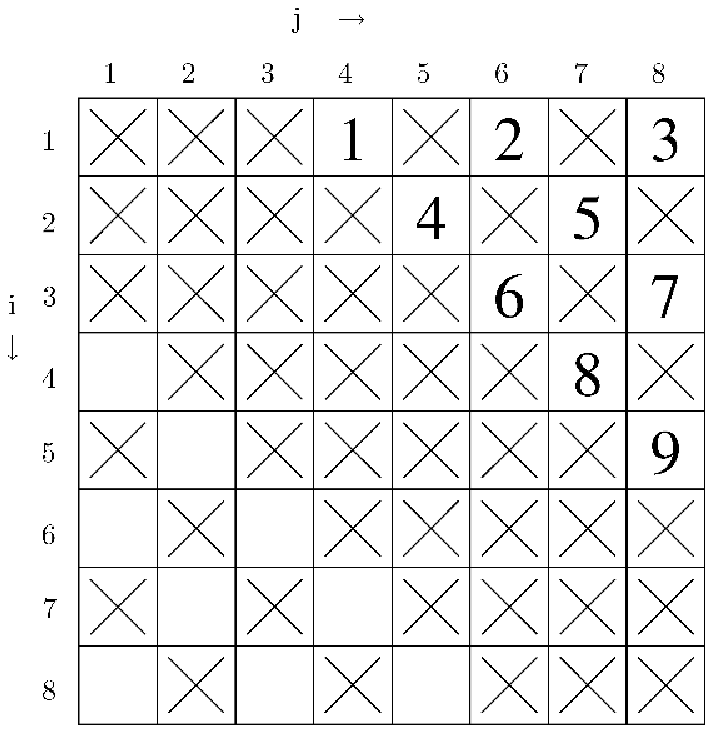}
 \end{center}
 \caption{Properties of contact maps of conformations consisting of $N=8$ monomers.
          Contacts that correspond to entries that are crossed out cannot be formed
          by conformations on the sc lattice. There are nine pairs of residues that
          can possibly be in contact (indexed fields).}
 \label{fig_cmprop}
\end{figure}
Figure~\ref{fig_cmprop} illustrates these properties for the $N=8$ case.
There are $Z_8=9$ possible contacts for conformations of that length,
i.e., nine bits are required to store any such contact map in memory.
In general, this number can be calculated to be
\begin{equation}
 \label{eqn_zn}
 Z_N =
 \begin{cases}
  \ \frac{1}{4}(N-2)^2 & \text{if } N \text{ even},\\
  \ \frac{1}{4}(N-3)(N-1) & \text{if } N \text{ odd.}
 \end{cases}
\end{equation}
For each contact map $C$, we also accumulate the number $g_\text{c}(C)$ of
self--avoiding walks corresponding to that contact map.
In the determination of $g_\text{c}(C)$
the trivial symmetries described above are automatically excluded. 
We include them in
a separate quantity $g_\text{s}(C)$ given by $\sum f_\text{s}(\mathbf{X})$,
where the sum runs over all $g_\text{c}(C)$ conformations that correspond to $C$
and $f_\text{s}(\mathbf{X})$ is given by~(\ref{eqn_fS}).
Knowledge of $g_\text{s}(C)$ is necessary for the calculation
of thermodynamic quantities and we store it for each contact map, too.
Furthermore, we retain for each contact map $C$ the last chain code
that we enumerate and whose conformation corresponds to $C$.
In particular, this yields all conformations corresponding
to contact matrices with $g_\text{c}(C)=1$ which allows to determine the
ground--state conformations of designing sequences.
\subsubsection{Parallelization}
The number of contact maps that can be simultaneously held in memory
was increased by distributing them over several individual processors (IPs).
We implemented the corresponding program according to the 
Message Passing Interface (MPI) standard\,\cite{Pach97} and executed it on 
the local Linux cluster
\emph{Hagrid}\footnote{\texttt{http://www.physik.uni-leipzig.de/Computer/Hagrid}},
consisting of 40 Athlon 1800+ MHz processors with 100 Mbit Ethernet communication.
\begin{figure}
 \includegraphics{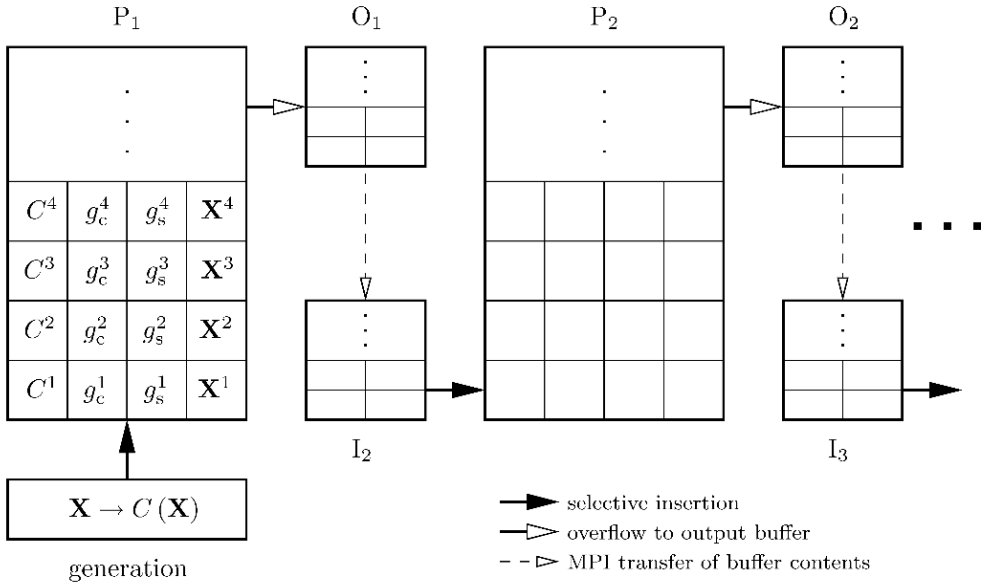}
 \caption{Generating contact maps in terms of parallel distributed programming.
          Two individual processors (IPs), P$_1$ and P$_2$, are shown. Each IP but P$_1$ disposes of an input
          buffer I$_i$, and there is an output buffer O$_i$ for all IPs but the last one.
          Buffering allows for faster information communication as less MPI messages
          have to be sent between the IPs.}
 \label{fig_parall}
\end{figure}
Figure~\ref{fig_parall} shows the basic structure of the program. The master
process P$_1$ generates all self--avoiding walks $\mathbf{X}$ and the corresponding
contact maps $C(\mathbf{X})$. For each self--avoiding walk, it can behave in three
different ways:
\begin{enumerate}
 \item
 If the memory associated to P$_1$ is not yet filled up and $C(\mathbf{X})$ was not
 stored
 in this memory partition before, $C(\mathbf{X})$ will be appended to the list of
 contact maps
 stored in P$_1$.
 \item
 If $C(\mathbf{X})$ is already stored in P$_1$ the corresponding counters $g_\text{c}(C)$
 and $g_\text{s}(C)$ will be increased by one and $f_\text{s}(\mathbf{X})$,
 respectively.
 \item
 If $C(\mathbf{X})$ is not stored in P$_1$ and its memory is completely filled,
 $C(\mathbf{X})$ will be stored in the master's output buffer O$_1$. When the
 output buffer is filled up all contact maps in the buffer will be transferred
 to the next IP's input buffer I$_2$.
\end{enumerate}
This way of storing contact maps is termed \emph{selective insertion} in Fig.~\ref{fig_parall}.
The behavior of the slave processes $\text{P}_2,\text{P}_3,\ldots$ is very similar.
The difference
is that their input buffers serve as sources of contact maps, they do not perform
any kind of enumeration. Their mere purpose is lookup and storage of contact maps.

The second step of the enumeration, i.e., going through all contact maps for all
sequences, can be trivially parallelized in order to reduce the running time. We achieved this
by simply distributing the set of contact maps over all IPs. Then, each IP
performs the enumeration with respect to its subset of contact maps. Finally,
all IPs send their enumeration results to the master process which compares the
lowest energies that were found by the slaves in order to find the ``globally''
minimal energies and the correct degeneracies $g_\text{c}$ and $g_\text{s}$. The
speed--up factor due to this parallelization is virtually equal to the number of
available processors.
\subsection{A simple example}
In the following, we illustrate briefly how the improvements discussed above
apply to the very simple $N=4$ example. There are $R_4=10$ relevant sequences
which we store in the array
\begin{equation}
 \textbf{S}=[\texttt{PPPP}, \texttt{PPPH}, \texttt{PPHP}, \texttt{PPHH}, \texttt{PHPH},
    \texttt{PHHP}, \texttt{PHHH}, \texttt{HPPH}, \texttt{HPHH}, \texttt{HHHH}].
 \label{eqn_seqN4}
\end{equation}
Also, there are six relevant conformations as shown in Fig.~\ref{fig_sawN4}. 
In the first enumeration step, all contact maps are determined.
Five conformations (\texttt{FFF}, \texttt{FFL}, \texttt{FLF}, \texttt{FLR}, and \texttt{FLU}) 
have no contacts and belong to the trivial empty contact
map which we will call $C^{(0)}$. The conformation encoded by \texttt{FLL} has
a single contact between its first and fourth residues; we refer to its
contact map as $C^{(1)}$. Thus, there are only two contact maps with 
$g_\text{c}(C^{(0)})=5$ and $g_\text{c}(C^{(1)})=1$, and
we compute $g_\text{s}(C^{(0)})=f_s(\texttt{FFF})+f_s(\texttt{FFL})+
f_s(\texttt{FLF})+f_s(\texttt{FLR})+f_s(\texttt{FLU})=126$ and $g_\text{s}(C^{(1)})=
f_s(\texttt{FLL})=24$.

In the second step, we run through all sequences from~(\ref{eqn_seqN4}) for
each of the two contact maps. The enumeration requires four more arrays
of length $R_4=10$ in order to
store the lowest energy, $\textbf{E}$, the accumulated degeneracies, $\textbf{G}_\text{c}$ and
$\textbf{G}_\text{s}$, and an example conformation for each sequence, $\textbf{W}$.
After evaluation of~(\ref{eqn_HP_Emain}) for all sequences with respect
to $C^{(0)}$ these arrays read
\begin{align}
 \textbf{E}&=[0,0,0,0,0,0,0,0,0,0],\\
 \textbf{G}_\text{c}&=[5,5,5,5,5,5,5,5,5,5],\\
 \textbf{G}_\text{s}&=[126,126,126,126,126,126,126,126,126,126],\\
 \textbf{W}&=[\texttt{FFF},\texttt{FFF},\texttt{FFF},\texttt{FFF},\texttt{FFF},\texttt{FFF},
     \texttt{FFF},\texttt{FFF},\texttt{FFF},\texttt{FFF}].\label{mbW}
\end{align}
Calculating now the energies with respect to $C^{(1)}$ yields once more
$E=0$ for the first seven sequences and a lower energy $E=-1$ for the last 
three sequences in (\ref{eqn_seqN4}). The arrays have to be updated accordingly. 
This means that for the sequences with energy $E=0$ the counter for the 
conformations, $\textbf{G}_\text{c}$, is incremented, while it is reset for the other
three sequences, since their energies are lower now. The degeneracy $\textbf{G}_\text{s}$
which includes the symmetry factors for the conformations is accumulated
appropriately and the new conformations possessing lower energies than those
in the previous step (\ref{mbW}) are stored in $\textbf{W}$:
\begin{align}
 \textbf{E}&=[0,0,0,0,0,0,0,-1,-1,-1],\label{eqn_EN4}\\
 \textbf{G}_\text{c}&=[6,6,6,6,6,6,6,1,1,1],\label{eqn_GCN4}\\
 \textbf{G}_\text{s}&=[150,150,150,150,150,150,150,24,24,24],\\
 \textbf{W}&=[\texttt{FFF},\texttt{FFF},\texttt{FFF},\texttt{FFF},\texttt{FFF},\texttt{FFF},
     \texttt{FFF},\texttt{FLL},\texttt{FLL},\texttt{FLL}].\label{eqn_SAWN4}
\end{align}
This shows that there are three designing sequences \texttt{HPPH}, \texttt{HPHH}, 
and \texttt{HHHH} corresponding to the entries equal to unity in~(\ref{eqn_GCN4}),
and from the corresponding entries in (\ref{eqn_SAWN4}) we read off
that, in this example, all three sequences possess the same unique 
ground--state conformation \texttt{FLL} with energy $E=-1$, as stored in~(\ref{eqn_EN4}).

Of course, parallelization must seem quite artificial in this very simple example. 
It would correspond to distributing the two contact maps $C^{(0)}$ and $C^{(1)}$
over two different IPs.
\section{Applications}
\label{sec_appl}
\begin{figure}
 \begin{center}
  \includegraphics[width=13cm]{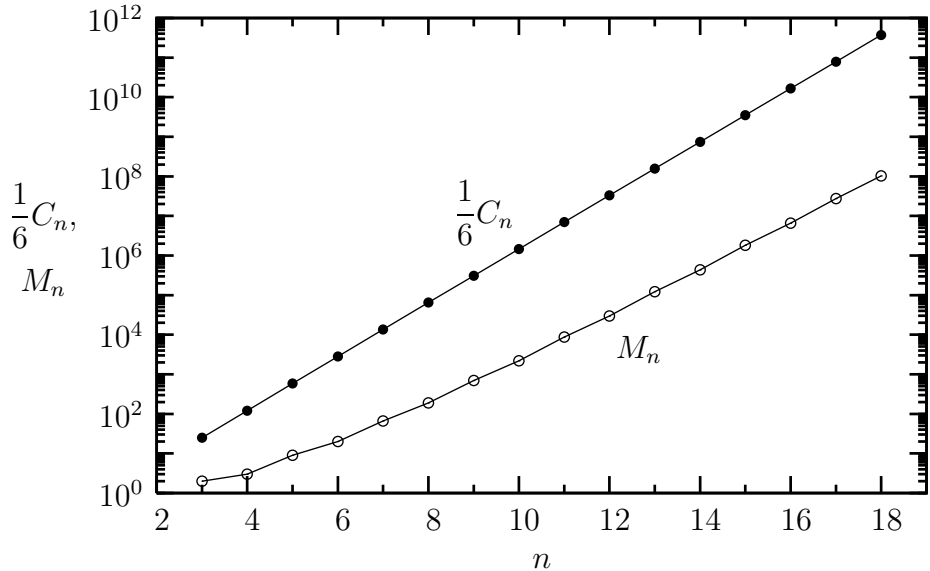}
 \end{center}
 \caption{Semi-log plot of the numbers $C_n$ of self--avoiding walks and
          numbers $M_n$ of contact maps vs.\
          the walk length $n=N-1$.}
 \label{fig_wm_pap}
\end{figure}

\begin{table}[b]
 \caption{Number of self-avoiding conformations $C_N$ and contact maps $M_N$
on a sc lattice.}
 \label{tab_wm_pap}
 \begin{center}
 \begin{tabular}{rrrrr}
  \\[-5mm]
  \hline
  \hline
  $N$ & $n=N-1$ & $\frac{1}{6}C_N$ & $M_N$ & $\frac{1}{6}C_N / M_N$\\
  \hline
  4 & 3 & 25 & 2 & 12.5\\[-2mm]
  5 & 4 & 121 & 3 & 40.3\\[-2mm]
  6 & 5 & 589 & 9 & 65.4\\[-2mm]
  7 & 6 & 2\,821 & 20 & 141.1\\[-2mm]
  8 & 7 & 13\,565 & 66 & 205.5\\[-2mm]
  9 & 8 & 64\,661 & 188 & 343.9\\[-2mm]
  10 & 9 & 308\,981 & 699 & 442.0\\[-2mm]
  11 & 10 & 1\,468\,313 & 2\,180 & 673.5\\[-2mm]
  12 & 11 & 6\,989\,025 & 8\,738 & 799.8\\[-2mm]
  13 & 12 & 33\,140\,457 & 29\,779 & 1\,112.9\\[-2mm]
  14 & 13 & 157\,329\,085 & 121\,872 & 1\,290.9\\[-2mm]
  15 & 14 & 744\,818\,613 & 434\,313 & 1\,714.9\\[-2mm]
  16 & 15 & 3\,529\,191\,009 & 1\,806\,495 & 1\,953.6\\[-2mm]
  17 & 16 & 16\,686\,979\,329 & 6\,601\,370 & 2\,527.8\\[-2mm]
  18 & 17 & 78\,955\,042\,017 & 27\,519\,000 & 2\,869.1\\[-2mm]
  19 & 18 & 372\,953\,947\,349 & 102\,111\,542 & 3\,652.4\\
  \hline
  \hline
 \end{tabular}
 \end{center}
\end{table}
Table~\ref{tab_wm_pap} and the corresponding Fig.~\ref{fig_wm_pap} show how
the number of self--avoiding walks, $C_N$, grows in comparison to the number of
contact maps, $M_N$, for chain lengths $N\leq 19$. 
For a given chain length, there are many more self--avoiding
walks than contact maps. The ratio between both numbers shown in the rightmost
column of Table~\ref{tab_wm_pap} keeps growing with $n=N-1$. 
The exponential growth, as suggested by Fig.~\ref{fig_wm_pap}, can generically be described by the following
scaling ansatz\,\cite{Gaun74,Gutt93}:
\begin{equation}
 C_n = A \mu^n n^{\gamma -1},
\label{mbScal}
\end{equation}
where $\gamma$ is
a universal exponent and $\mu$ the effective coordination number.
For self--avoiding walks
we reproduce the well--known results $\mu_\text{SAW} \approx 4.684$ and $\gamma\approx 1.16$~\cite{MacD92,MacD00,Chen02}
by means of a ratio method analysis (see, e.g., Refs.~\cite{Gaun74,Stan87,Gutt89}).
Assuming a scaling form (\ref{mbScal}) also for the number of contact maps (CM), a similar analysis yields
$\mu_\text{CM}\approx 4.38$, i.e., their (still) exponential growth is slower than that
of self--avoiding walks (see Ref.~\cite{Schi04} for more details).

In the second enumeration step we determined all designing sequences of
length $N\leq 19$, their numbers are shown in Table~\ref{tab_des}.
As the interaction is more
complicated in the MHP case, it is intuitively clear that degeneracies are
lifted and that there are hence more designing
sequences for that model than in the HP case. We also note that there are
fewer designing sequences in the HP model on the sc lattice than for the
same model and the same lengths $N$ on the square lattice\,\cite{Irb02}. 
\begin{table}[h]
 \caption{Numbers of designing sequences $S_N$ (only relevant sequences, see text) in the HP and MHP models.}
 \label{tab_des}
 \begin{tabular}{r|rrrrrrrrrrrrrrrr}
  \multicolumn{17}{c}{}\\[-5mm]
  \hline
  \hline
  $N$&$\!$4&$\!$5&$\!$6&$\!$7&$\!$8&$\!$9&$\!$10&$\!$11&$\!$12&$\!$13&$\!$14&$\!$15&
  $\!$16&$\!$17&$\!$18&$\!$19\\
  \hline
  $S^{\text{HP}}_N$&$\!$3&$\!$0&$\!$0&$\!$0&$\!$2&$\!$0&$\!$0&$\!$0&$\!$2&$\!$0&$\!$1&$\!$1&
  $\!$1&$\!$8&$\!$29&$\!$47\\
  $S^{\text{MHP}}_N$&$\!$7&$\!$0&$\!$0&$\!$6&$\!$13&$\!$0&$\!$11&$\!$8&$\!$124&$\!$14&$\!$66&
  $\!$97&$\!$486&$\!$2\,196&$\!$9\,491&$\!$4\,885\\
  \hline
  \hline
 \end{tabular}
\end{table}
\section{Summary and Outlook}
\label{sec_sum}
In the first part of our exact enumeration procedure we generated the
complete sets of contact maps for self--avoiding walks of $n\leq 18$
steps, i.e.\ for conformations of up to $N=19$ monomers. We parallelized
our program such as to distribute the set of contact maps over several
memory partitions of a Linux cluster.
In the second step of enumeration, we determined all designing
sequences for both types of interactions considered. Here, parallelization
is used to decrease the required computer time.

The results obtained this way can be used in a statistical analysis of
designing sequences and their ground--state conformations. First, in the
space of sequences, we can discuss the \emph{hydrophobicity}, i.e., the
H--content of designing sequences as well as \emph{hydrophobicity profiles}
describing the distribution of hydrophobic monomers in the polymer chain.
Additionally, it enables us to investigate in how far monomers are involved in the formation of
$HH$ contacts (and $HP$ contacts in case of the MHP model) by defining
\emph{hydrophobic contact density profiles}. Second, in the space of conformations,
the data obtained herein allow for the study of the \emph{end--to--end distances}
and \emph{radii of gyration} as measures of the compactness of designed
conformations. The consideration of the distribution
of the \emph{designability} of designed conformations shows that some
conformations are preferred over others as ground--state conformations
of designing sequences. The complete statistical analysis can be
found in Ref.~\cite{Schi04}.

Finally, it should be pointed out that a slight
variation of the enumeration procedure explained herein also allows for the
\emph{exact} determination of the density of states $g(E)$, i.e., the
number of conformations corresponding to all energy levels and not just to
the ground--state energy. This number includes all symmetries which
is why we store the degeneracies $g_\text{s}(C)$ for all contact maps $C$
(see Section~\ref{ssec_cm}). For a given HP sequence, $g(E)$ can be used
to determine the temperature dependence of energetic quantities, in
particular that of the specific heat whose peaks can be associated with conformational
transitions~\cite{Schi04}.
\section{Acknowledgments}
This work was partially supported by the German--Israel--Foundation (GIF)
under Grant No.\ I--653--181.14/1999. One of us (R.S.) acknowledges
support by the Studienstiftung des deutschen Volkes.


\begin{thebibliography}{99}
 \bibitem{Crei93} T.~E.~Creighton, \emph{Proteins: Structures and Molecular Properties},
                  2nd ed. (W.~H.~Freeman and Company, New York, 1993).
 \bibitem{Dill85} K.~A.~Dill, Biochemistry \textbf{24}, 1501 (1985);
                  K.~F.~Lau and K.~A.~Dill, Macromolecules \textbf{22}, 3986 (1989).
 \bibitem{Tang00} C.~Tang, Physica A \textbf{288}, 31 (2000).
 \bibitem{Miya96} S.~Miyazawa and R.~L.~Jernigan, J.~Mol.~Biol.~\textbf{256}, 623 (1996).
 \bibitem{Soka95} A.~D.~Sokal, \emph{Monte Carlo Methods for the Self--Avoiding Walk},
                  in \emph{Monte Carlo and Molecular Dynamics Simulations in Polymer
                  Science}, edited by K.~Binder (Oxford University Press, New York, 1995),
                  p.~51.
 \bibitem{Ung93}  R.~Unger and J.~Moult, J.~Mol.~Biol.~\textbf{231}, 75 (1993).
 \bibitem{Okam98} Y.~Okamoto, Recent Research Developments in Pure \& Applied Chemistry
                  \textbf{2}, 1 (1998).
 \bibitem{Born97} E.~Bornberg--Bauer, \emph{Chain Growth Algorithms for HP--Type Lattice
                  Proteins}, in {\it Proceedings of the First International Conference on
                  Computational Molecular Biology}, Santa Fe (ACM Press, New York, 1997), p.~47.
 \bibitem{Gras97} P.~Grassberger, Phys.~Rev.~E \textbf{56}, 3682 (1997).
 \bibitem{Mit01}  A.~Mitsutake, Y.~Sugita, and Y.~Okamoto, Biopolymers (Peptide Science)
                  \textbf{60}, 96 (2001).
 \bibitem{Bach03} M.~Bachmann and W.~Janke, Phys.~Rev.~Lett.~\textbf{91}, 208105 (2003);
                  J.~Chem.~Phys.~\textbf{120}, 6779 (2004).
 \bibitem{Schi04} R.~Schiemann, M.~Bachmann, and W.~Janke, q-bio.BM/0405009, to appear
   in J.\ Chem.\ Phys.\ (in print).
 \bibitem{Irb02}  A.~Irb\"ack and C.~Troein, J.~Biol.~Phys. \textbf{28}, 1 (2002).
 \bibitem{Li96}   H.~Li, R.~Helling, C.~Tang, and N.~Wingreen, Science \textbf{273}, 666 (1996).
 \bibitem{Cejt02} H.~Cejtin, J.~Edler, A.~Gottlieb, R.~Helling, H.~Li, J.~Philbin,
                  N.~Wingreen, and C.~Tang, J.~Chem.~Phys.~\textbf{116}, 352 (2002).
 \bibitem{MacD92} D.~MacDonald, D.~L.~Hunter, K.~Kelly, and N.~Jan, J. Phys. A: Math. Gen.
                  \textbf{25}, 1429 (1992).
 \bibitem{MacD00} D.~MacDonald, S.~Joseph, D.~L.~Hunter, L.~L.~Moseley, N.~Jan, and
                  A.~J.~Guttmann, J. Phys. A: Math. Gen. \textbf{33}, 5973 (2000).
 \bibitem{Chen02} M.~Chen and K.~Y.~Lin, J. Phys. A: Math. Gen. \textbf{35}, 1501 (2002).
 \bibitem{Pach97} P.~S.~Pacheco, \emph{Parallel Programming with MPI}, 1st printed ed.
                  (Morgan Kaufmann, San Francisco, 1997).
 \bibitem{Gaun74} D.~S.~Gaunt and A.~J.~Guttmann, \emph{Asymptotic Analysis of Coefficients},
                  in \emph{Phase Transitions and Critical Phenomena}, Vol.~3, edited by C.~Domb
                  and M.~S.~Green (Academic Press, London, 1974), p.~181.
 \bibitem{Gutt93} J.~L.~Guttmann and A.~J.~Guttmann, J.~Phys.~A: Math.~Gen.~\textbf{26}, 
                  2485 (1993).
 \bibitem{Stan87} H.~E.~Stanley, \emph{Introduction to Phase Transitions and Critical
                  Phenomena} (Oxford University Press, New York, 1987).
 \bibitem{Gutt89} A.~J.~Guttmann, \emph{Asymptotic Analysis of Power--Series Expansions},
                  in \emph{Phase Transitions and Critical Phenomena}, Vol.~13, edited by
                  C.~Domb and J.~L.~Lebowitz (Academic Press, London, 1989), p.~3.
\end{thebibliography}
\end{document}